\documentclass[conference]{IEEEtran}
\IEEEoverridecommandlockouts

\usepackage{cite}
\usepackage{url}
\usepackage{amsmath}
\usepackage{graphicx}
\usepackage{amssymb}
\usepackage{xcolor}
\usepackage{url}
\usepackage{array}
\usepackage{subfig}

\graphicspath{{Figures/}}

\begin{document}

\title{Multi-Year Spectral Structure of 6G Candidate Bands at 2.7~GHz and 4.4~GHz\thanks{This work was supported in part by the National Science Foundation under Grants CNS-2332835 and CNS-2450593.}
}

\author{
\IEEEauthorblockN{Amir Hossein Fahim Raouf}
\IEEEauthorblockA{
\textit{Department of Electrical and Computer Engineering} \\
\textit{North Carolina State University} \\
Raleigh, NC, USA \\
amirh.fraouf@ieee.org}
\and
\IEEEauthorblockN{\.{I}smail G\"{u}ven\c{c}}
\IEEEauthorblockA{
\textit{Department of Electrical and Computer Engineering} \\
\textit{North Carolina State University} \\
Raleigh, NC, USA \\
iguvenc@ncsu.edu}
}
\maketitle

\begin{abstract}
Mid-band spectrum between 2 and 8 GHz is a critical resource for sixth-generation (6G) systems as it uniquely balances favorable propagation characteristics with scalable bandwidth. Recent U.S. policy highlights candidate bands near 2.7, 4.4, and 7.1 GHz, all of which host substantial federal and non-federal incumbency, including high-power radiolocation and aeronautical telemetry systems. Although these segments are being considered for potential relocation of federal incumbents to enable commercial use, their long-term viability depends on the structural integrity of the spectrum. In such environments, the practical value of spectrum depends on the reliability and contiguity of available spectrum opportunities.
This paper presents a measurement-driven feasibility analysis of two representative segments, 2.69--2.9~GHz and 4.4--4.94~GHz, using Software-Defined Radio~(SDR) measurements collected during Packapalooza campaigns from 2022 to 2025. Deployment-oriented metrics are introduced to quantify scan-window reliability (SWR), altitude-dependent usable spectrum availability ratio (USAR), largest contiguous clean bandwidth (LCCB), spectral fragmentation, and extreme interference excursions.
The results reveal significant year-to-year structural variability. In the 2.69--2.9~GHz band, USAR remains near unity in 2022 and 2023, but drops to approximately 0.65 in 2024 and 0.8 in 2025, accompanied by fragmentation and limited contiguous bandwidth across altitudes. The 4.4--4.94~GHz band exhibits a similar temporal pattern, but with smaller reliability degradation and larger contiguous support, often exceeding several hundred megahertz even during incumbent-dominant periods. The results highlight that wideband feasibility in these candidate bands depends strongly on spectral contiguity and structural stability rather than nominal bandwidth alone.
\end{abstract}

\section{Introduction}

Mid-band spectrum between 2~GHz and 8~GHz plays a central role in the evolution of sixth-generation (6G) wireless systems. Compared with millimeter-wave allocations, these frequencies provide a favorable compromise among coverage, penetration, and scalable channel bandwidth, enabling wideband air interfaces while maintaining practical link budgets~\cite{ITUIMT2030}. International harmonization efforts in the upper 6~GHz range target contiguous allocations of several hundred megahertz to support high-throughput carrier configurations and flexible numerologies~\cite{CEPT6GHz}.

In the United States, however, several candidate mid-band segments under consideration for future commercial use are heavily occupied by federal and non-federal incumbents, including 2.69--2.9~GHz, 4.4--4.94~GHz, and portions of 7.125--7.4~GHz~\cite{WH_6G_Memo_2025}. These bands host radiolocation systems, aeronautical telemetry, fixed microwave links, and satellite services operating with substantial transmit power and structured emission patterns. Table~\ref{tab:incumbents_summary} summarizes the principal incumbent services and their typical interference characteristics.

\begin{table*}[t] 
\centering 
\caption{Primary U.S. Incumbent Services and Characteristics in Candidate 6G Mid-Band Segments \cite{NTIA_NSS_2023, WH_6G_Memo_2025}.} \label{tab:incumbents_summary} 
\begin{tabular}{|>{\raggedright\arraybackslash}p{1.8cm}|>{\raggedright\arraybackslash}p{4.6cm}|>{\raggedright\arraybackslash}p{4cm}|>{\raggedright\arraybackslash}p{6cm}|} 
\hline 
\textbf{Band Segment} & \textbf{Primary Federal Services} & \textbf{Current Non-Federal Status} & \textbf{Typical Interference Characteristics} \\ 
\hline 
2.69--2.9 GHz & Radiolocation (ASR-9/11, NEXRAD), Aeronautical/Maritime radionavigation \cite{NTIA_NSS_2023, FCC_47CFR_2_106} & Radio Astronomy (2.69--2.7 GHz), Secondary Experimental \cite{FCC_47CFR_2_106} & High-power pulsed radar emissions, rotating beams, localized high-peak excursions, sensitive passive receivers in adjacent 2.69 GHz \\ 
\hline 
4.4--4.94 GHz & Aeronautical Mobile Telemetry (AMT), Tactical Military systems \cite{NTIA_NSS_2023, NTIA_AMT_2022} & 4.94--4.99 GHz National Band Manager framework (Public Safety), International 5G/6G candidate \cite{ITUR_WRC23_Final} & Frequency-selective occupancy, high-EIRP airborne emitters (AMT), geometry-dependent coupling, altitude-sensitive interference profiles \\ 
\hline 
7.125--8.4 GHz & Federal Fixed Service, Fixed-Satellite (uplink), Space Research, EESS \cite{NTIA_NSS_2023, FCC_47CFR_2_106} & 7.125--7.4 GHz prioritized for commercial reallocation, overlap with unlicensed U-NII-5/8 \cite{WH_6G_Memo_2025} & Narrowband point-to-point links, high-gain satellite uplinks, aggregate interference from densified terrestrial 6G and unlicensed deployments \\ 
\hline 
\end{tabular} 
\end{table*}

In such incumbent-dense environments, nominal bandwidth alone does not determine operational deployability~\cite{flinck2025overview, raouf2026beyond}. We define operational deployability as the ability of a band to sustain reliable, contiguous wideband operation at the relevant reconfiguration time scale under incumbent coexistence constraints. In practice, deployability depends on three interrelated structural properties: 1) scan-window reliability~(SWR) at the operational time scale, 2) spectral contiguity sufficient for wideband carrier formation, and 3) bounded extreme interference excursions that do not exceed radio frequency~(RF) front-end dynamic-range limits. Together, these properties determine whether a nominally wide allocation can support stable, high-throughput operation in shared-spectrum conditions.

Existing spectrum characterization approaches are often insufficient for this purpose. Measurement campaigns frequently report average occupancy or duty-cycle statistics~\cite{WellensOccupancy}. While valuable for regulatory planning, such metrics do not reveal whether usable spectrum forms sufficiently contiguous blocks, persists over decision intervals relevant to coordination mechanisms, or remains robust to rare but severe interference peaks. Short-horizon analyses based on real-world mid-band measurements show that channel availability varies across minute-scale time horizons and that average occupancy metrics do not adequately capture the structural behavior relevant to dynamic spectrum access~\cite{mao2026ai}. Furthermore, high-power interference events directly affect User Equipment~(UE) blocking and adjacent-channel selectivity~\cite{3GPP38_101_1}. As a result, extreme interference levels become practical receiver design constraints rather than merely statistical outliers.

These observations motivate a deployment-oriented structural evaluation framework for incumbent-heavy mid-band spectrum. This paper contributes a measurement-based methodology that quantifies scan-window reliability, largest contiguous clean bandwidth, spectral fragmentation, and extreme interference excursions using multi-year Software-Defined Radio~(SDR) measurements collected during four annual urban campaigns. By linking spectrum identification with practical wideband carrier formation under coexistence constraints, the proposed framework provides an empirical evidence for assessing the feasibility of dynamic 6G deployment in incumbent-dense environments. 

\looseness=-1
It is important to note that the incumbent landscape in these bands is subject to ongoing policy review and potential reallocation. Federal systems may be relocated, reconfigured, or transitioned over time. The analysis in this paper therefore evaluates structural deployability under the presently observed interference environment, without presuming permanent incumbency.

\section{Related Work}

The strategic importance of mid-band spectrum for future International Mobile Telecommunications~(IMT) systems is articulated in multiple 6G vision and framework documents, including ITU IMT-2030 and 3GPP study items on 6G scenarios and requirements~\cite{ITUIMT2030, 3GPP38914, NextGAlliance}. These documents emphasize the need for wide, contiguous allocations capable of supporting scalable numerologies and high-throughput air interfaces. In particular, harmonization initiatives in the upper 6~GHz range target channel bandwidths on the order of several hundred megahertz to enable wideband carrier configurations and simplified RF design~\cite{CEPT6GHz}. However, such discussions primarily address spectrum identification and high-level performance objectives, with limited empirical analysis of deployability under incumbent coexistence.

Dynamic spectrum access in incumbent-heavy bands has been studied extensively, most prominently in the 3.5~GHz Citizens Broadband Radio Service~(CBRS) framework~\cite{sohul2015spectrum}. Related radar–communications coexistence research has developed protection criteria, exclusion zones, interference modeling approaches, and coordination mechanisms~\cite{zheng2019radar}. While these works provide important regulatory and architectural foundations, they focus largely on coexistence enforcement and protection requirements rather than on whether sufficiently large and structurally stable wideband opportunities emerge within the shared spectrum.

Measurement-driven spectrum characterization has further examined utilization across time and geography~\cite{WellensOccupancy}. Such campaigns typically report average occupancy, duty cycle, or power statistics. Although valuable for regulatory planning, these metrics do not fully capture structural properties that determine wideband feasibility. Specifically, average utilization does not indicate whether usable frequencies form contiguous blocks, whether availability persists over operational decision intervals, or whether rare high-power excursions impose front-end blocking or dynamic-range constraints. Wideband 6G air interfaces require coherence across frequency and stability over time, not merely low mean occupancy.

More recent 6G discussions highlight artificial intelligence~(AI)-native spectrum management, adaptive sharing, and fine-grained coordination in heterogeneous environments~\cite{NextGAlliance, 3GPP38914}. In such settings, structural characteristics including contiguity, fragmentation, and temporal reliability become central performance determinants, as they directly constrain achievable bandwidth, aggregation strategies, and scheduling flexibility. The present work addresses this gap by integrating multi-year measurement data with deployment-oriented structural metrics, thereby providing an empirical framework that links spectrum identification with the practical formation of wideband carriers under contiguity, reliability, and RF blocking constraints in incumbent-dense mid-band environments.

\section{Structural Characterization Metrics}
\label{sec:metrics}

This section defines the measurement-driven metrics used to evaluate the operational deployability of 6G systems under incumbent coexistence. Rather than relying on aggregate utilization statistics, these metrics quantify sustained usability, spectral contiguity, and interference morphology as observed in multi-year measurement data. These dimensions map directly to 6G performance requirements across three primary technical domains:

\begin{itemize}
    \item \textbf{Service Availability:} The reliability and temporal persistence of spectrum opportunities at operational scales determine the feasibility of sustained wideband links.
    
    \item \textbf{Peak and User-Experienced Data Rates:} The availability of contiguous usable bandwidth imposes fundamental bounds on achievable throughput and the practical configuration of wideband waveforms.
    
    \item \textbf{RF Design and Energy Efficiency:} High-power interference excursions constrain receiver headroom, thereby impacting RF front-end design margins and overall system energy efficiency~\cite{ITUIMT2030, 3GPP38914}.
\end{itemize}

Mapping these high-level 6G performance objectives to concrete, actionable metrics requires a high-fidelity characterization of the spectral environment across multiple spatial and temporal dimensions. The subsequent subsections briefly review the experimental methodology and develop the formal mathematical framework utilized to derive structural characterization metrics from the empirical spectral observations.

\subsection{Measurement Campaign and Dataset}

The metrics developed herein are derived from the AERPAW spectrum-monitoring campaigns conducted during the annual Packapalooza events at North Carolina State University. The sensing platform, deployed on a tethered Helikite, captured sub-6~GHz spectrum activity in a dense urban environment under realistic network load conditions.

While the experimental architecture and data collection methodologies are detailed in~\cite{maeng2025altitude, raouf2025wireless}, a brief overview is provided for context. The payload utilized a USRP-based Software Defined Radio (SDR) integrated with a GPS receiver. The system operated in a passive sensing mode, sweeping center frequencies from 87~MHz to 6~GHz. Although the Helikite altitude was controlled via a ground tether, aerodynamic drift and operational constraints resulted in non-uniform altitude trajectories across different campaigns, necessitating the altitude-binning approach described in Section~\ref{subsec:reliability}.

The multi-year datasets, made publicly available through the Dryad repository~\cite{hossein2024packapalooza2024, maeng2023packapalooza2023, maeng2023packapalooza, hossein2024aerpaw, maeng2023aerpaw, raouf2025aerpaw}, provide calibrated power spectra with synchronized timestamps and altitude metadata. Let $P(f,t,h)$ denote the measured aggregated receiver power in frequency bin $f$ at time $t$ and altitude $h$, where $P(\cdot)$ represents the calibrated SDR output with a frequency resolution of $\Delta f = 60$~kHz. Given that absolute calibration and the effective noise floor may exhibit variance across campaigns, usability is defined relative to a conservative, band-specific noise reference derived from the measurement data.

\subsection{Measurement Time Resolution and Scan-Window}
\label{subsec:time_windows}
\looseness = -1
The dataset provides per-frequency timestamps and altitudes, so $t$ and $h$ may vary across frequency bins within a sweep. In the Packapalooza campaigns considered here, one sweep spans approximately 15~s, implying that sub-second incumbent structure is not directly resolved at the sweep level. We therefore aggregate samples into fixed-duration \emph{scan-window} of length $\Delta t = 60$~s, aligned with the implementation.

All temporal reliability statements in this paper are interpreted at the $\Delta t$ time scale. Incumbent signals with sub-second dynamics (e.g., rotating radars with short dwell time) may appear intermittent or persistent depending on scan rate, revisit time, and the aggregation induced by $\Delta t$. Accordingly, these results should be interpreted as feasibility constraints for spectrum access methods operating on second-to-minute reconfiguration time scales (e.g., database-assisted selection or modest-latency channel changes), not as a substitute for sub-second sensing.

\subsection{Noise Reference and Usability Threshold}
\label{subsec:threshold}

For each year and band, we estimate a conservative band noise reference $N_{\mathrm{band}}$ using a two-stage robust procedure consistent with the code. First, for each frequency bin, we compute a lower temporal quantile (10th percentile across time) to suppress intermittent interferers and approximate the underlying noise-dominated level. Second, we apply a frequency-domain quantile (25th percentile across bins) to mitigate bias from bins that are persistently occupied. The result is a scalar $N_{\mathrm{band}}$ that serves as a conservative operational reference for that band-year.

The usability threshold is then defined as
\begin{equation}
T_{6G} = N_{\mathrm{band}} + \Delta,
\end{equation}
where $\Delta = 10$~dB is a deployment margin representing a minimum SNR headroom target under practical link budgets.

At the sample level, a measurement is considered \emph{usable} if
\begin{equation}
P(f,t,h) < T_{6G}.
\end{equation}
This binary usability indicator is the input to the scan-window reliability metrics below.

\subsection{Scan-Window Usability and Reliability}
\label{subsec:reliability}

To characterize the reliability of candidate 6G mid-band segments, the framework evaluates usability at the scan-window level. This approach accounts for cases where multiple samples occupy a single time, frequency, and altitude coordinate $(f,h)$ during a measurement interval. Let $\mathcal{W}_{f,h,k}$ denote the set of samples within the $k$-th scan-window of duration $\Delta t$ at frequency $f$ and altitude $h$. The within-window usable fraction is defined as
\begin{equation}
\eta_{f,h,k} =
\frac{1}{|\mathcal{W}_{f,h,k}|}
\sum_{t \in \mathcal{W}_{f,h,k}}
\mathbf{1}\{P(f,t,h) < T_{6G}\},
\end{equation}
where $T_{6G}$ represents the coexistence threshold. A scan-window is considered usable if it satisfies the reliability condition
\begin{equation}
\eta_{f,h,k} \ge 1 - \epsilon.
\end{equation}
In this study, $\epsilon = 0.05$ is utilized to enforce a 95\% within-window cleanliness requirement.

Let $\mathcal{K}_{f,h}$ represent the set of scan-windows possessing sufficient sample support. The altitude-dependent frequency reliability, $p_{\mathrm{usable}}(f,h)$, is defined as the fraction of these supported windows that meet the reliability constraint:
\begin{equation}
p_{\mathrm{usable}}(f,h)
=
\frac{1}{|\mathcal{K}_{f,h}|}
\sum_{k \in \mathcal{K}_{f,h}}
\mathbf{1}\{\eta_{f,h,k} \ge 1-\epsilon\}.
\end{equation}
This reliability primitive serves as the foundation for the structural metrics detailed in the following sections, including the Usable Spectrum Availability Ratio (USAR), contiguity, and spectral fragmentation.

\noindent \textbf{Implementation Notes:}
\looseness=-1
The analysis framework enforces threshold constraints on both the sample-level support per altitude--time--frequency coordinate and the aggregate scan-window reliability per $(f,h)$ pair. This ensures that the reported $p_{\mathrm{usable}}(f,h)$ values are derived from a statistically sufficient number of temporal snapshots captured throughout the experiment period. To prevent isolated frequency bin volatility, which often arises from the snapshot nature of the receiver sweeps, a five-bin moving-average filter is applied to the reliability mask. All subsequent structural metrics are calculated based on this smoothed scan-window reliability state.

\subsection{USAR}
\label{subsec:usar}

Let $\mathcal{B}$ denote the set of frequency bins in the band, and $N_b = |\mathcal{B}|$. A frequency bin is considered \emph{reliably usable} at altitude $h$ if $p_{\mathrm{usable}}(f,h) \ge 1-\epsilon$. The Usable Spectrum Availability Ratio is defined as
\begin{equation}
\mathrm{USAR}(h) =
\frac{1}{N_b}
\sum_{f \in \mathcal{B}}
\mathbf{1}\{p_{\mathrm{usable}}(f,h) \ge 1-\epsilon\}.
\end{equation}
USAR measures the fraction of the band that is reliably usable at altitude $h$ under the scan-window reliability constraint. In Key Performance Indicator~(KPI) terms, USAR is best interpreted similar to an \emph{availability} indicator of how much spectrum can be depended upon at the operational time scale $\Delta t$; it does not by itself guarantee wideband feasibility, which depends on contiguity.

\subsection{Contiguity}
\label{subsec:lccb}

Wideband air interfaces require contiguous frequency support to form large carriers without excessive carrier aggregation overhead. Let
\[
\mathcal{C}(h) = \{f \in \mathcal{B} : p_{\mathrm{usable}}(f,h) \ge 1-\epsilon\}
\]
denote the reliably usable bins at altitude $h$. The Largest Contiguous Clean Bandwidth~(LCCB) is defined as
\begin{equation}
\mathrm{LCCB}(h)
=
\max_{\mathcal{S} \subseteq \mathcal{C}(h)}
|\mathcal{S}| \Delta f,
\end{equation}
where $\mathcal{S}$ spans contiguous bins and $\Delta f = 60$~kHz.
LCCB provides a direct structural proxy for the maximum single-carrier bandwidth supportable under coexistence at altitude $h$. In KPI terms, it constrains the \emph{upper bound} on achievable throughput from a single contiguous allocation and indicates whether wideband numerologies are practically supportable without heavy aggregation.

\subsection{Fragmentation}
\label{subsec:sfi}

Let $\{L_i(h)\}$ denote the lengths (in bins) of contiguous segments within $\mathcal{C}(h)$. The Spectral Fragmentation Index~(SFI) is defined as
\begin{equation}
\mathrm{SFI}(h)
=
1 - \frac{\max_i L_i(h)}{\sum_i L_i(h)}.
\end{equation}
SFI approaches zero when usable spectrum forms a single dominant contiguous block and increases as it fragments into multiple smaller segments. Fragmentation matters operationally because it can increase aggregation complexity, guard-band overhead, and control-plane scheduling burden. In KPI terms, higher SFI can indirectly degrade energy efficiency and achievable user rate through increased overhead, even when USAR remains moderate.

\subsection{Maximum Measured Power}
\label{subsec:power_maps}

For completeness, we also report altitude--frequency maps of the maximum measured aggregated power:
\begin{equation}
P_{\max}(f,h) = \max_t P(f,t,h),
\end{equation}
evaluated over all available samples at each $(f,h)$.
$P_{\max}(f,h)$ is useful for characterizing rare high-power excursions that can drive receiver headroom requirements. In practice, such excursions translate directly into RF implementation constraints, including front-end compression risk and required analog-to-digital converter~(ADC) dynamic-range margin. Rare high-power events can therefore determine receiver back-off and blocking tolerance requirements, even when average interference levels remain moderate.

\section{Measurement Results}
\label{sec:results}

We evaluate two mid-band segments representative of candidate U.S. 6G allocations: 2.69--2.9~GHz and 4.4--4.94~GHz. The measurements were collected using the AERPAW helikite-based spectrum monitoring platform during multiple annual campaigns conducted at the Packapalooza festival (urban) and the Lake Wheeler site (rural) in the Raleigh, NC, USA area. Details of the measurement campaign and associated dataset are provided in~\cite{maeng2025altitude, raouf2025wireless}.

For each band and year, the usability threshold $T_{6G}$ is derived independently using the two-stage quantile procedure described in Section~\ref{subsec:threshold}, with $\Delta=10$~dB and $\epsilon=0.05$. Reliability is evaluated using the scan-window formulation at $\Delta t=60$~s granularity (Section~\ref{subsec:time_windows}). Altitude samples are grouped into 10~m bins to improve statistical stability and visualization clarity.
Certain altitude bins do not appear in the results when the minimum data requirements are not satisfied. Specifically, frequency–altitude cells are retained only if they satisfy minimum sample requirements in both the time and frequency dimensions (at least two samples per dimension in this paper). Consequently, gaps in altitude coverage reflect insufficient data support rather than structural spectrum effects.

All statements regarding reliability and usability are therefore tied explicitly to the 60~s operational decision interval~(i.e., SWR). Sub-second incumbent dynamics, such as rotating radar scan cycles, are not resolved at this granularity and are outside the scope of the present analysis.

\subsection{2.69--2.9~GHz}
\label{subsec:results_2p7}

\begin{figure}[t]
\centering
\includegraphics[width=0.87\linewidth]{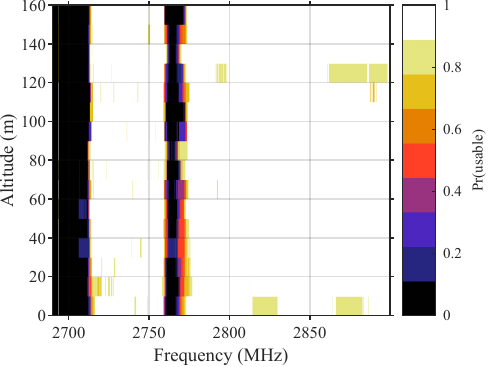}
\caption{Altitude--frequency map of $p_{\mathrm{usable}}(f,h)$ for 2.69--2.9~GHz band at Packapalooza (urban), 2025.}
\label{fig:2025_2p7_2p9GHz_pusable_fh}
\end{figure}

Fig.~\ref{fig:2025_2p7_2p9GHz_pusable_fh} shows that the dominant reliability impairment in 2025 is strongly frequency selective. Specifically, $p_{\mathrm{usable}}(f,h)$ is persistently low near the lower band edge (approximately 2.69--2.71~GHz) across essentially all altitudes, and a narrow reliability notch appears around 2.76--2.77~GHz. Outside these regions, most of the band exhibits near-unity reliability over a wide altitude range. The altitude dependence is comparatively weak relative to the frequency dependence, suggesting incumbent activity concentrated in specific sub-bands rather than a band-wide elevation effect.

\begin{figure}[t]
\centering

\subfloat[USAR versus altitude.]{
\includegraphics[width=0.48\linewidth]{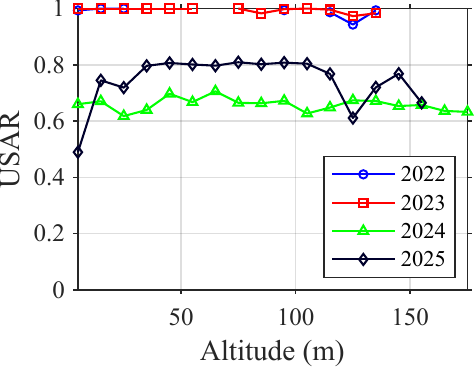}
\label{fig:multidataset_2p7_2p9GHz_usar}
}
\subfloat[LCCB versus altitude.]{
\includegraphics[width=0.48\linewidth]{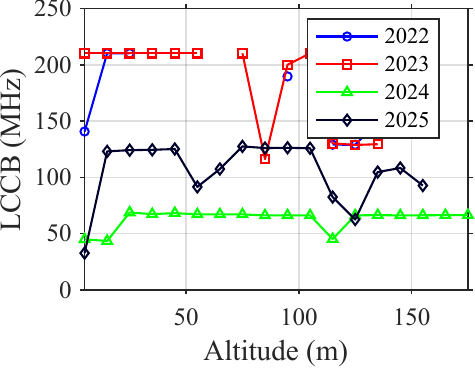}
\label{fig:multidataset_2p7_2p9GHz_lccb}
}

\vspace{-0.3cm}

\subfloat[SFI versus altitude.]{
\includegraphics[width=0.48\linewidth]{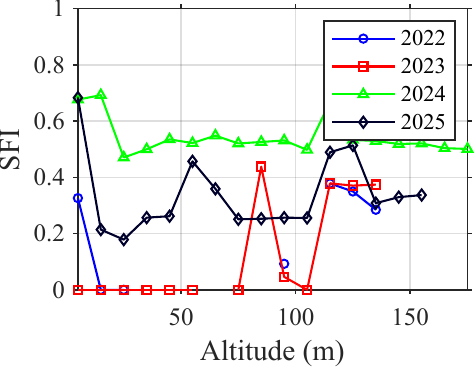}
\label{fig:multidataset_2p7_2p9GHz_sfi}
}

\caption{Altitude-dependent structural metrics for the 2.69--2.9~GHz band across multiple urban datasets (2022--2025): \textbf{(a)} USAR, \textbf{(b)} LCCB, and \textbf{(c)} SFI.}
\label{fig:multidataset_2p7_2p9GHz_metrics}

\end{figure}


The year-to-year impact of these frequency-localized impairments is captured by USAR in Fig.~\ref{fig:multidataset_2p7_2p9GHz_usar}. In both 2022 and 2023, USAR is essentially unity across most altitudes, indicating that nearly the full segment satisfies the 60~s reliability criterion. In 2024, USAR drops and remains approximately 0.6--0.7 across altitudes, reflecting persistent loss of reliably usable bins. In 2025, USAR is generally higher than 2024 (around 0.8 over much of the altitude range) but exhibits a pronounced dip near 120--130~m, indicating that reliability deficits can couple to geometry at particular heights even when the primary impairment is frequency localized.


Contiguity provides a more direct indicator of wideband feasibility than aggregate availability. Fig.~\ref{fig:multidataset_2p7_2p9GHz_lccb} shows that in 2024 the largest contiguous clean block is constrained to roughly 60--70~MHz over most altitudes. In 2025, LCCB is larger on average (typically around 125~MHz at many altitudes) but collapses near the same altitude interval where USAR dips, reaching on the order of 62~MHz around 120--130~m. At the lowest altitude bin, 2025 also shows reduced contiguity (tens of MHz), consistent with localized strong interference in the power maxima. In contrast, 2022 and 2023 measurements support substantially larger contiguous blocks (often approaching the full 210~MHz segment), consistent with their near-unity USAR.

\looseness = -1
Fig.~\ref{fig:multidataset_2p7_2p9GHz_sfi} shows fragmentation trends that are consistent with the LCCB contiguity results. The 2024 campaign exhibits sustained fragmentation (SFI $\approx$0.5 across most altitudes), consistent with usable spectrum being split into multiple comparable segments. In 2025, fragmentation is moderate and altitude dependent (typically 0.2--0.45, rising near the altitude interval where LCCB collapses). By contrast, 2022--2023 show near-zero SFI over broad altitude ranges, indicating a dominant contiguous usable block.

\begin{figure}[t]
\centering
\includegraphics[width=0.87\linewidth]{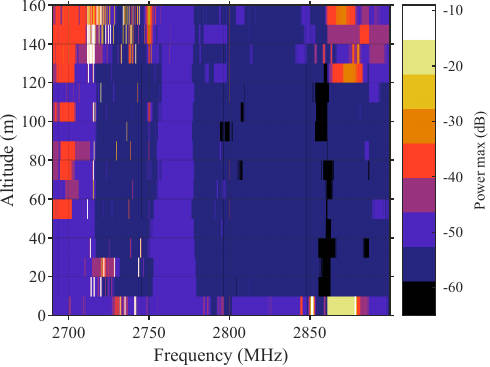}
\caption{Maximum measured aggregated power $P_{\max}(f,h)$ for 2.69--2.9~GHz band at Packapalooza (urban), 2025.}
\label{fig:2025_2p7_2p9GHz_powerMax_fh}
\end{figure}

The maximum-power map in Fig.~\ref{fig:2025_2p7_2p9GHz_powerMax_fh} provides practical RF insight: elevated maxima are concentrated near the same frequency regions that exhibit reduced $p_{\mathrm{usable}}$, particularly near the lower band edge and parts of the upper edge. This indicates that infeasibility is driven not only by frequent exceedances of $T_{6G}$ (captured by $p_{\mathrm{usable}}$) but also by intermittent high-power peaks that impose receiver dynamic-range requirements.

The structural limitations observed in the 2.69--2.9 GHz segment (persistent frequency-localized reliability deficits, moderate USAR in several measurement years, and limited contiguous support) highlight the challenges for wideband single-carrier deployment under incumbent coexistence at the 60~s decision scale. These characteristics suggest that, without changes to incumbent usage, the band may not readily support broad unlicensed or full-power commercial deployment as currently configured. In light of recent U.S. policy directing formal study of this band's suitability for commercial 6G use and potential relocation of federal systems if feasible, such empirical structural insight can inform those assessments by clarifying whether incumbency patterns are ``evenly distributed and manageable,'' or whether relocation or intensive sharing mechanisms would be required to achieve large contiguous blocks suitable for wideband carriers.

\begin{figure}[!t]
        \centering
        \subfloat[Lake Wheeler (rural) in 2022]{ \includegraphics[width=0.48\linewidth]{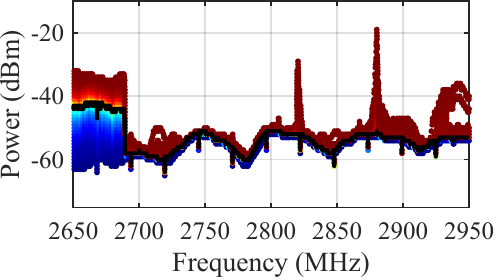}\label{fig:Fig_Wheeler_2022_FreqPowerCDF_n77_2650_2950MHz_2022}}
        \subfloat[Lake Wheeler (rural) in 2024]{ \includegraphics[width=0.48\linewidth]{Figures/Fig_Wheeler_2022_FreqPowerCDF_n77_2650_2950MHz_2022.pdf}\label{fig:Fig_Wheeler_2022_FreqPowerCDF_n77_2650_2950MHz_2022}}\\
         \subfloat[Packapalooza (urban) in 2022]{ \includegraphics[width=0.48\linewidth]{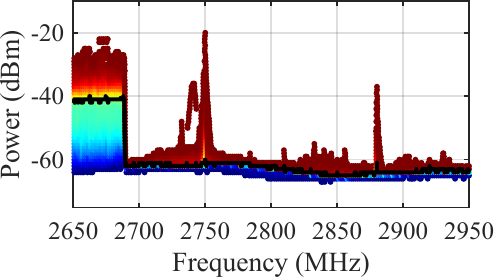}\label{fig:Fig_Pack_2022_FreqPowerCDF_n77_2650_2950MHz_2022}}
        \subfloat[Packapalooza (urban) in 2023]{ \includegraphics[width=0.48\linewidth]{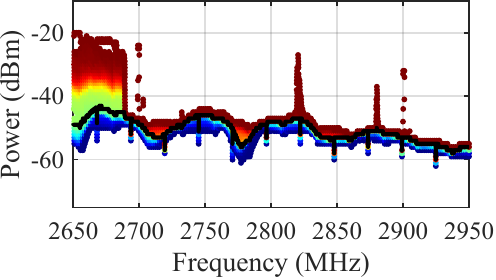}\label{fig:Fig_Pack_2022_FreqPowerCDF_n77_2650_2950MHz_2022}}\\
        \subfloat[Packapalooza (urban) in 2024]{ \includegraphics[width=0.48\linewidth]{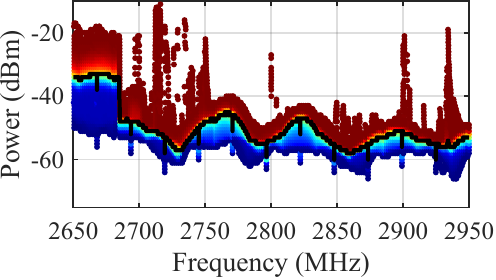}\label{fig:Fig_Pack_2024_FreqPowerCDF_n77_2650_2950MHz_2024}}
        \subfloat[Packapalooza (urban) in 2025]{ \includegraphics[width=0.48\linewidth]{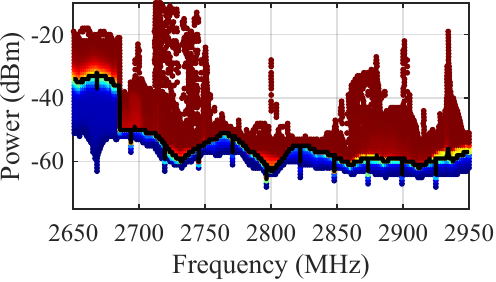}\label{fig:Fig_Pack_2025_FreqPowerCDF_n77_2650_2950MHz_2025}}
	\caption{CDF and median of received power across 2650--2950~MHz derived from helikite measurements. The black curve shows the median power across snapshots, while colored markers indicate empirical CDF values for each frequency bin. \textbf{(a)} Wheeler 2022, \textbf{(b)} Wheeler 2024, \textbf{(c)} Packapalooza 2022, \textbf{(d)} Packapalooza 2023, \textbf{(e)} Packapalooza 2024, \textbf{(f)} Packapalooza 2025.}\label{fig:spec_cdf}
\end{figure}

Fig.~\ref{fig:spec_cdf} presents the empirical power distribution across 2650–2950~MHz derived from helikite measurements in both rural (Lake Wheeler) and urban (Packapalooza) environments over multiple years. While the primary candidate band of interest is 2690–2900~MHz, an additional 50~MHz margin is included on each side to capture activity in adjacent bands and reveal potential guard-band or leakage effects near the band edges. The colored markers represent the empirical CDF of received power for each frequency bin, while the black curve indicates the median power across measurement snapshots. The lower portion of the extended range (approximately 2650–2690~MHz) exhibits similarly elevated power levels in both environments, indicating persistent incumbent activity. Differences become more visible in the central portion of the band, where the urban Packapalooza measurements show a larger spread of power values and more frequent high-power peaks at varying frequencies. These peaks are not consistently aligned across years, suggesting that they arise from temporally varying transmissions and changing channel utilization rather than persistent fixed-frequency emitters.

\begin{figure}[t]
\centering
\includegraphics[width=0.87\linewidth]{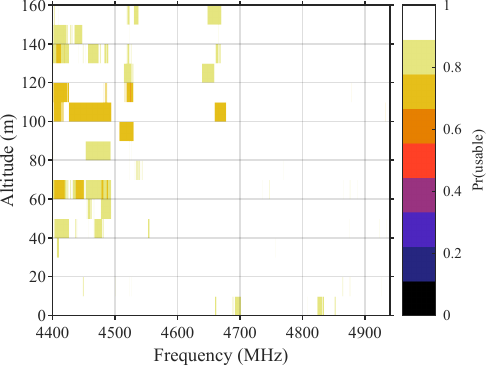}
\caption{Altitude--frequency map of $p_{\mathrm{usable}}(f,h)$ for 4.4--4.94~GHz band at Packapalooza (urban), 2025.}
\label{fig:2025_4p4_4p94GHz_pusable_fh}
\end{figure}

\subsection{4.4--4.94~GHz}
\label{subsec:results_4p4}

The 4.4--4.94~GHz reliability map in Fig.~\ref{fig:2025_4p4_4p94GHz_pusable_fh} is largely high across the band, but the remaining impairment is not uniformly distributed. Reduced $p_{\mathrm{usable}}$ appears as patchy, localized regions, concentrated primarily in the lower portion of the segment (roughly 4.40--4.52~GHz) and most evident over mid-to-high altitudes. Relative to 2.69--2.9~GHz, the reliability loss is less pervasive and does not partition the band into a small number of consistently unusable sub-bands.

\begin{figure}[t]
\centering

\subfloat[USAR versus altitude.]{
\includegraphics[width=0.48\linewidth]{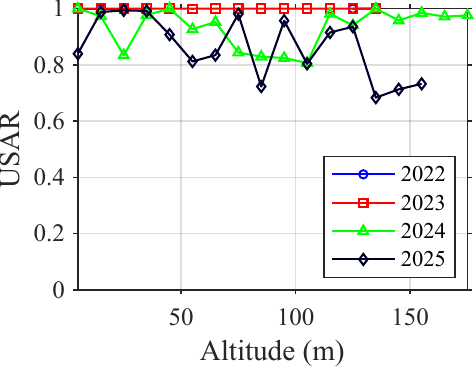}
\label{fig:multidataset_4p4_4p94GHz_usar}
}
\subfloat[LCCB versus altitude.]{
\includegraphics[width=0.48\linewidth]{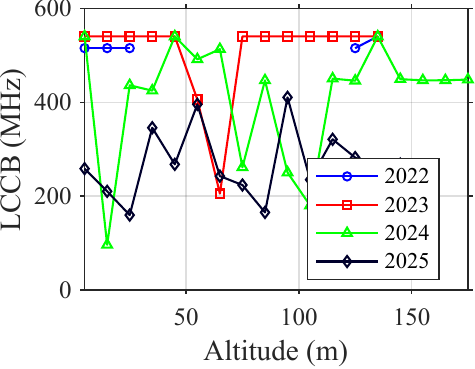}
\label{fig:multidataset_4p4_4p94GHz_lccb}
}

\vspace{-0.3cm}

\subfloat[SFI versus altitude.]{
\includegraphics[width=0.48\linewidth]{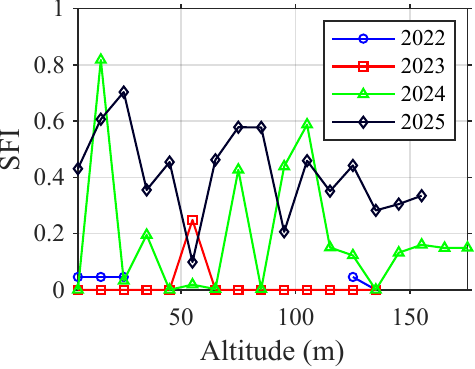}
\label{fig:multidataset_4p4_4p94GHz_sfi}
}

\caption{Altitude-dependent structural metrics for the 2.4--4.94~GHz band across multiple urban datasets (2022--2025). \textbf{(a)} USAR, \textbf{(b)} LCCB, and \textbf{(c)} SFI.}
\label{fig:multidataset_2p7_2p9GHz_metrics}

\end{figure}


USAR (Fig.~\ref{fig:multidataset_4p4_4p94GHz_usar}) remains high in 2022--2023 (near unity over most altitudes). In 2024, USAR is generally between about 0.8 and 1.0 with a mid-altitude dip. In 2025, USAR is more variable, remaining relatively high at low-to-mid altitudes but degrading more clearly above 130~m, where it drops to approximately 0.68. Thus, a majority of the segment remains usable at the 60~s decision scale across all years, although altitude-dependent degradation becomes noticeable at higher elevations.


High aggregate reliability often translates into large contiguous support, but the year dependence remains important. As shown in Fig.~\ref{fig:multidataset_4p4_4p94GHz_lccb}, 2022--2023 frequently support contiguous blocks exceeding 500~MHz, approaching the full segment. In 2024, contiguity is still substantial but exhibits some altitude-dependent drops. In 2025, LCCB is consistently lower than 2022--2023 and highly non-monotone with altitude, varying roughly between 160 and 410~MHz depending on height. This indicates that wideband single-carrier operation is often feasible in this segment; however, the maximum supportable carrier bandwidth can vary significantly with altitude.


\looseness = -1
Fig.~\ref{fig:multidataset_4p4_4p94GHz_sfi} shows fragmentation trends that are consistent with the observed LCCB variations. In 2022--2023, SFI stays near zero across most altitudes, indicating a dominant contiguous usable block. In 2024 and especially 2025, SFI increases and becomes altitude dependent, implying that reliability losses occur through partial band break-up rather than wholesale band-wide blocking. Importantly, even when SFI rises, the LCCB curves show that a dominant contiguous region often persists.

\begin{figure}[t]
\centering
\includegraphics[width=0.87\linewidth]{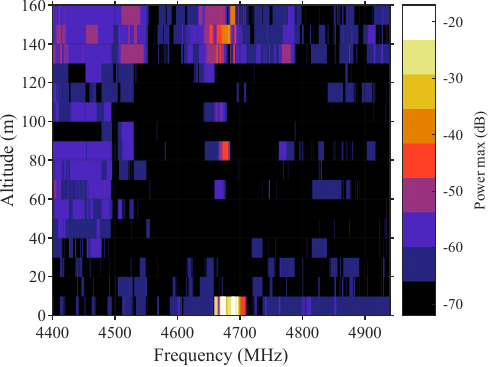}
\caption{Maximum measured aggregated power $P_{\max}(f,h)$ for 4.4--4.94~GHz band at Packapalooza (urban), 2025.}
\label{fig:2025_4p4_4p94GHz_powerMax_fh}
\end{figure}

The maximum-power map in Fig.~\ref{fig:2025_4p4_4p94GHz_powerMax_fh} shows that high-power excursions are spatially and spectrally localized, with prominent peaks around the mid-band (near 4.66--4.71~GHz) at low altitude and additional elevated regions at higher altitudes. This supports a practical interpretation: compared to 2.69--2.9~GHz, impairment is less dominated by a small number of persistently unusable sub-bands, but extreme excursions can still impose RF margin requirements and contribute to altitude-dependent contiguity drops.

This segment is generally more favorable for forming wide contiguous carriers at the 60~s decision scale. In 2024--2025, reliability degradation is smaller than in 2.69--2.9~GHz, but is more clearly altitude coupled at higher elevations, leading to non-trivial variations in LCCB and moderate fragmentation.

\begin{figure}[!t]
        \centering
        \subfloat[Lake Wheeler (rural) in 2022]{ \includegraphics[width=0.48\linewidth]{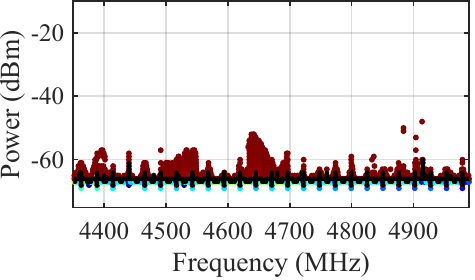}\label{fig:Fig_Wheeler_2022_FreqPowerCDF_4_4GHz_2022}}
        \subfloat[Lake Wheeler (rural) in 2024]{ \includegraphics[width=0.48\linewidth]{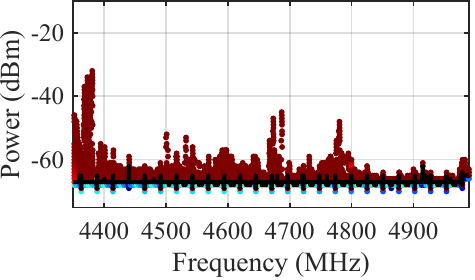}\label{fig:Fig_Wheeler_2024_FreqPowerCDF_4_4GHz_2024}}\\
         \subfloat[Packapalooza (urban) in 2022]{ \includegraphics[width=0.48\linewidth]{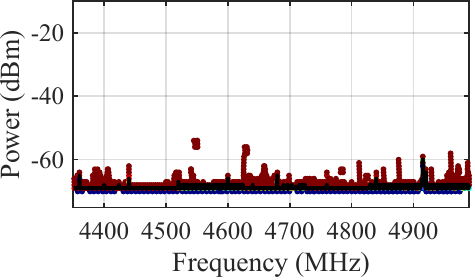}\label{fig:Fig_Pack_2022_FreqPowerCDF_4_4GHz_2022}}
        \subfloat[Packapalooza (urban) in 2023]{ \includegraphics[width=0.48\linewidth]{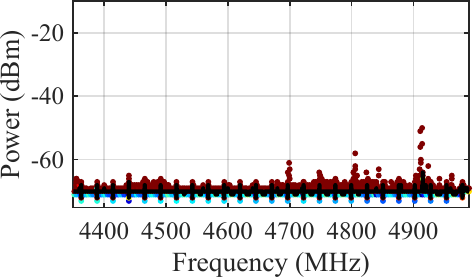}\label{fig:Fig_Pack_2023_FreqPowerCDF_4_4GHz_2023}}\\
        \subfloat[Packapalooza (urban) in 2024]{ \includegraphics[width=0.48\linewidth]{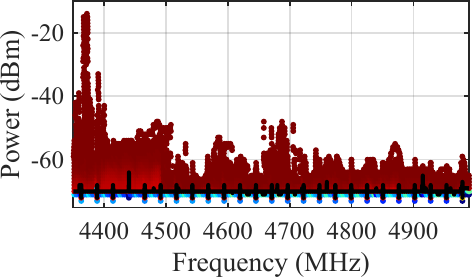}\label{fig:Fig_Pack_2024_FreqPowerCDF_4_4GHz_2024}}
        \subfloat[Packapalooza (urban) in 2025]{ \includegraphics[width=0.48\linewidth]{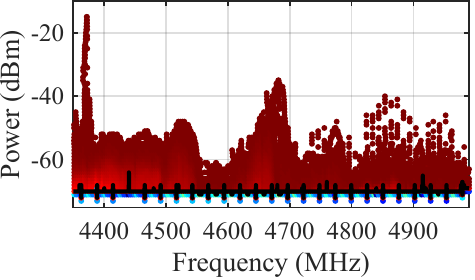}\label{fig:Fig_Pack_2025_FreqPowerCDF_4_4GHz_2025}}
	\caption{CDF and median of received power across 4350--4990~MHz derived from helikite measurements. The black curve shows the median power across snapshots, while colored markers indicate empirical CDF values for each frequency bin. \textbf{(a)} Wheeler 2022, \textbf{(b)} Wheeler 2024, \textbf{(c)} Packapalooza 2022, \textbf{(d)} Packapalooza 2023, \textbf{(e)} Packapalooza 2024, \textbf{(f)} Packapalooza 2025.}\label{fig:spec_cdf_lw_4_4}
\end{figure}

\looseness = -1
Fig.~\ref{fig:spec_cdf_lw_4_4} presents the empirical power distribution across 4350–4990~MHz derived from helikite measurements at Lake Wheeler (rural) and Packapalooza (urban) over multiple years. Across all datasets, the median received power remains close to the measurement floor, indicating the absence of persistent wideband activity in this band. The colored CDF markers reveal sporadic high-power peaks at scattered frequencies; however, their occurrence and locations vary substantially between years. Some datasets exhibit clusters of peaks at particular frequencies, while others remain largely near the background level. The lack of consistent frequency alignment across years suggests that these events arise from intermittent transmissions rather than stable incumbents occupying fixed channels.

\section{Conclusion and Forward-Looking Discussion}

This study evaluated the structural feasibility of two incumbent-heavy mid-band segments currently discussed in the U.S. 6G pipeline. Using deployment-oriented metrics derived from multi-year measurement data, we quantified temporal reliability, largest contiguous clean bandwidth, spectral fragmentation, and extreme interference behavior at a 60~s operational decision scale.
The results demonstrate that deployability is governed by spectral structure rather than nominal bandwidth. The 4.4--4.94~GHz segment exhibits consistently high reliability and substantial contiguous clean regions across most altitudes and campaigns, indicating structural conditions compatible with wideband operation under dynamic access frameworks. In contrast, the 2.69--2.9~GHz segment shows persistent frequency-selective reliability constraints and reduced contiguous support in several measurement years. In this regime, wideband feasibility becomes sensitive to channel placement, aggregation strategy, and receiver dynamic-range margin.

Two interpretive considerations are important.
First, feasibility is explicitly indexed to time scale. All reliability metrics are defined at $\Delta t = 60$~s granularity. The conclusions therefore inform coordination mechanisms operating on second-to-minute time scales, such as database-assisted access or periodic channel reassignment. Sub-second incumbent dynamics are averaged within this interval and may require complementary high-temporal-resolution campaigns if waveform adaptation is expected to track scan-cycle behavior.

Second, extreme-interference statistics represent implementation constraints rather than abstract tail descriptors. Expressing interference excursions relative to $T_{6G}$ directly links measurement results to receiver blocking tolerance, adjacent-channel selectivity, and dynamic-range requirements. In incumbent-heavy environments, rare high-power events can dominate front-end design even when average occupancy remains moderate.

From a spectrum-policy perspective, the measurements establish a structural baseline against which future band evolution can be assessed. Partial clearing through relocation would be expected to increase contiguous clean bandwidth in affected sub-bands, whereas structured sharing without full clearing places greater emphasis on intrinsic spectral coherence and receiver robustness. Changes in incumbent composition or adjacent-band utilization may alter fragmentation patterns even without formal reallocation.
The central finding is therefore conditional but engineering-driven: as candidate mid-band segments evolve from clearing toward coordination, practical 6G feasibility depends on whether the resulting spectrum opportunities remain (i) temporally reliable at the relevant reconfiguration time scale, (ii) sufficiently contiguous to support wideband carriers, and (iii) bounded in extreme interference excursions to avoid prohibitive receiver margin requirements.

\bibliographystyle{IEEEtran}
\bibliography{refs}

\end{document}